\documentclass[namedreferences]{solarphysics}
\usepackage[optionalrh]{spr-sola-addons} % For Solar Physics 
\usepackage{graphicx}        % For eps figures, newer & more powerfull
\usepackage{amssymb}        % useful mathematical symbols
\usepackage{color}           % For color text: \color command
\usepackage{url}             % For breaking URLs easily trough lines
\usepackage[pdfborder={0 0 0 },urlcolor=blue,breaklinks]{hyperref}
\ifx \arxivurl  \undefined \def \arxivurl#1{\href{http://arxiv.org/abs/#1}{\textsf{#1}}}\fi %% JWL <<<<<<<<<<<<<<<<<<<<<<<<<<<<<<<<<<<<<<
\ifx \doiurl    \undefined \def \doiurl#1{\href{http://dx.doi.org/#1}{\textsf{#1}}}\fi %% JWL <<<<<<<<<<<<<<<<<<<<<<<<<<<<<<<<<<<<<<
\ifx \adsurl    \undefined \def \adsurl#1{\href{http://adsabs.harvard.edu/abs/#1}{\textsf{#1}}}\fi %% JWL <<<<<<<<<<<<<<<<<<<<<<<<<<<<<<<<<<<<<<

\newcommand{\adv}{    {\it Adv. Space Res.}}

\newcommand{\apj}{    {\it Astrophys. J.}}

\newcommand{\mnras}{  {\it Mon. Not. Roy. Astron. Soc.}}

\newcommand{\solphys}{{\it Solar Phys.}}

\newcommand{\ssr}{    {\it Space Sci. Rev.}}

\begin{document}

\begin{article}

\begin{opening}

\title{An alternative measure of solar activity from detailed sunspot datasets}

\author{J.~\surname{Murak\"ozy}\sep
        T.~\surname{Baranyi}\sep
        A.~\surname{Ludm\'any}
       }

\runningauthor{J. Murak\"ozy \emph{et al.}}
\runningtitle{Solar Activity Measure}

   \institute{Heliophysical Observatory, Research Centre for Astronomy and Earth Sciences, Hungarian Academy of Sciences,
\\4010 Debrecen, P.O. Box 30, Hungary \\email: \url{murakozy.judit@csfk.mta.hu} 
email: \url{baranyi.tunde@csfk.mta.hu} email: \url{ludmany.andras@csfk.mta.hu}}

\begin{abstract}
The sunspot number is analyzed by using detailed sunspot data, including aspects of observability, sunspot sizes, and proper identification of sunspot groups as discrete entities of the solar activity. The tests show that besides the subjective factors there are also objective causes of the ambiguities in the series of sunspot numbers. To introduce an alternative activity measure the physical meaning of the sunspot number has to be reconsidered. It contains two components whose numbers are governed by different physical mechanisms, this is one source of the ambiguity. This article suggests an activity index, which is the amount of emerged magnetic flux. The only long-term proxy measure is the detailed sunspot area dataset with proper calibration to the magnetic flux amount. The Debrecen sunspot databases provide an appropriate source for the establishment of the suggested activity index.

\end{abstract}
\end{opening}
%-------------------------------------------------

\section{Introduction}
           \label{S-Introduction}

It is a common problem of all long-term studies in astronomy and space physics that their empirical backgrounds, the observed datasets are necessarily inhomogeneous because of the large variety of data sources and uneven observational coverage. The causes of the inhomogeneities are various. During the centuries several circumstances may be modified substantially, e.g. the atmospheric seeing, instrumental background (mainly optics), and registration technique (visual, graphic, photographic, electronic). Subjective causes of the inhomogeneity may be the differences between the members of generations of observers, their personal biases and also mistakes.

The Sunspot Number ($R_{z}$), the most important, indispensable long-term parameter of the level of solar activity, was introduced by Rudolf Wolf in 1848 at the Z\"urich Eidgen\"ossische Sternwarte. He reconstructed the $R_{z}$ by collecting and studying  all earlier observations made after 1610. The next directors of the observatory, Alfred Wolfer, Wilhelm Brunner and Max Waldmeier continued his work. Although the observing instrument, the strategy and seeing were more or less identical, the perfect long term homogeneity could not be guaranteed. This dataset is currently continued by the SILSO team at the Royal Observatory, Brussels \cite{Clette07}.

\inlinecite{Clette14} conducted a thorough analysis of the homogeneity problems of the sunspot number and identified some causes of the discontinuities in the long term variations. They have carried out the necessary corrections to improve the homogeneity of the datasets of the sunspot number and the group sunspot number \cite{Hoyt} as well as the variation of their ratio. The corrected dataset provides an improved view of the long term variation of solar activity.

The present work has two aims: on the one hand testing the classic definition of the sunspot number with detailed datasets, in particular with sunspot area data, on the other hand testing a proposed new solar activity parameter based on detailed sunspot data.

\section {Problems with the Wolf number}
             \label{S-Wolf}

\subsection {Observational problems}
             \label{S-Wolfobs}

\begin{figure}    
   \centerline{\includegraphics[width=0.5\textwidth,clip=,angle=-90]{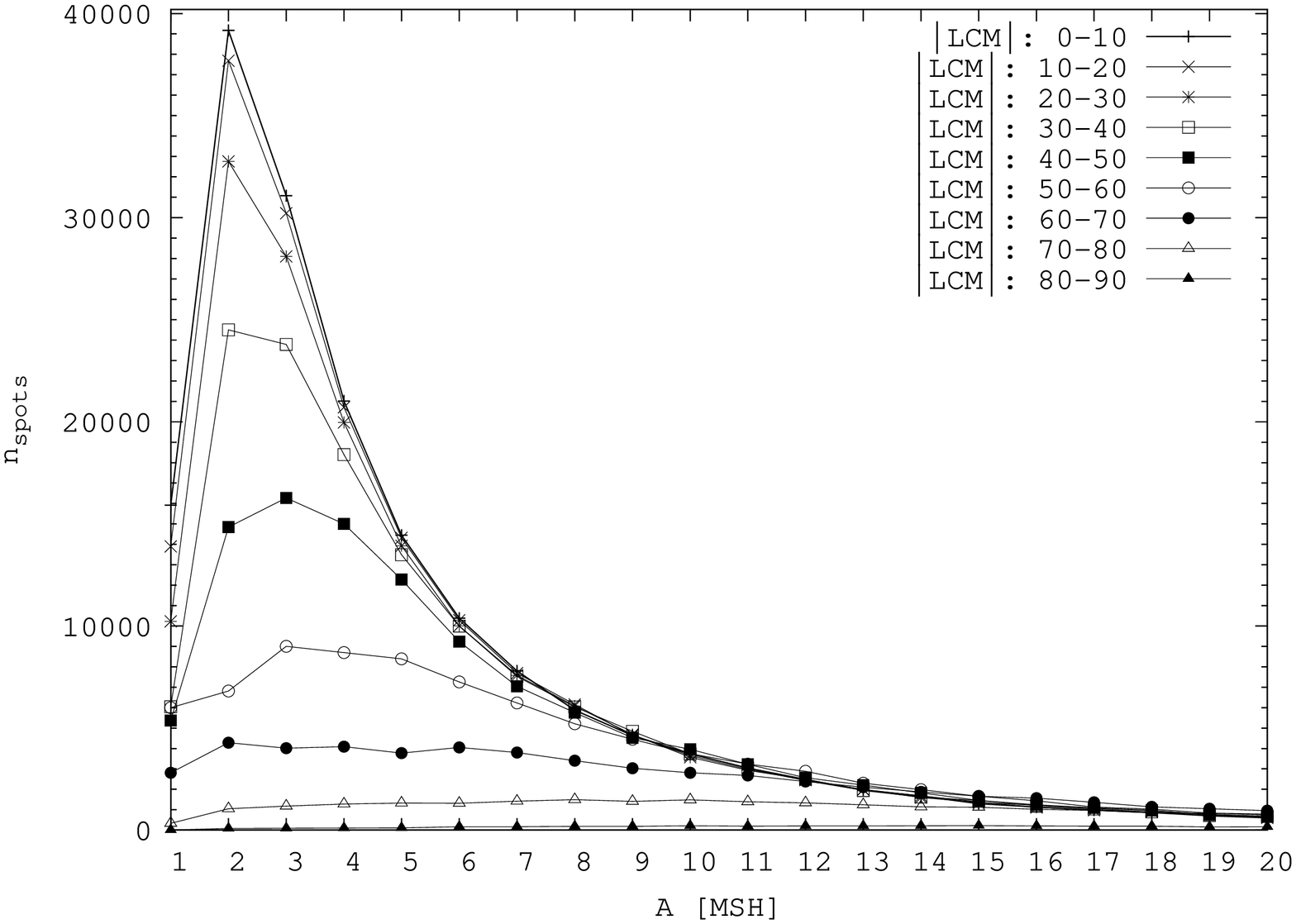}}
   \centerline{\includegraphics[width=0.5\textwidth,clip=,angle=-90]{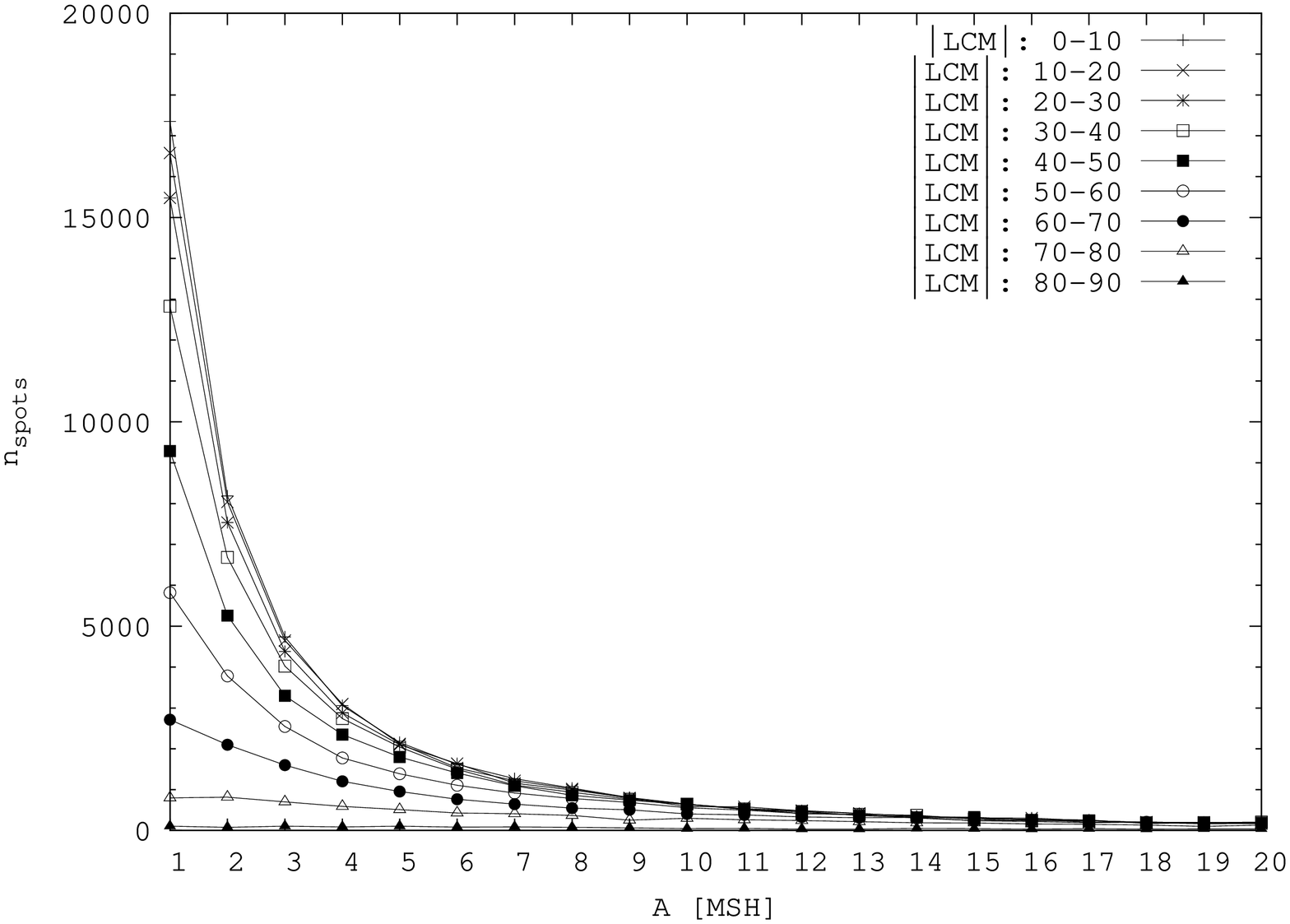}}

   \caption {Umbral area distribution at different distances from the solar disc center derived from SDD data (upper panel) and DPD data (lower panel).}

   \label{LCM_area}
   \end{figure}

The tremendous importance of the sunspot number in long term studies is unquestionable but some limitations may be worth mentioning. One of the critical issues in determining the sunspot number is the observability of sunspots. This does not only mean a pure technical or seeing problem that could be eliminated by averaging the inputs of many sources, it has a principal constraint: the center-limb variation of the observability of  sunspots. Figure~\ref{LCM_area} shows the numbers of sunspots of different areas detected in longitudinal bins of 10 degrees at increasing longitudinal distances from the Central Meridian (LCM). 

It can be seen that the numbers of observed spots strongly depend on both their sizes and distances from the central region. The two diagrams also demonstrate that the spatial resolution of the instruments has a non negligible impact on the result, the SDD data (SOHO/MDI Debrecen Data, left panel) are based on SOHO/MDI observations whose resolution is low to detect areas of 1 MSH (Millionth of Solar Hemisphere) so the maximum is at 2 MSH. The inputs to the DPD (Debrecen Photoheliographic Data) are the Debrecen/Gyula ground based observations, these have higher resolution, thus the weight of 1 MSH spots is larger in this sample. However, the center-limb variation is a more serious problem than a simple instrumental limit, it means that the chance of observing a sunspot is in principle lower close to the limb than close to the center. This difference decreases in the cases of larger spots but it is not neglectable even for areas of 5 MSH. This is not a subjective restriction.

This LCM-dependence imposes the same constraint on each input dataset of ISSN, it cannot cause hidden jumps. However, it makes questionable, whether the daily sunspot number is a real measure of the activity. It may have a significant daily variability even if all spots remain the same during two weeks, just because of the variable observability.  The monthly values are presumably real.

\begin{figure}    
   \centerline{\includegraphics[width=0.70\textwidth,clip=,angle=-90]{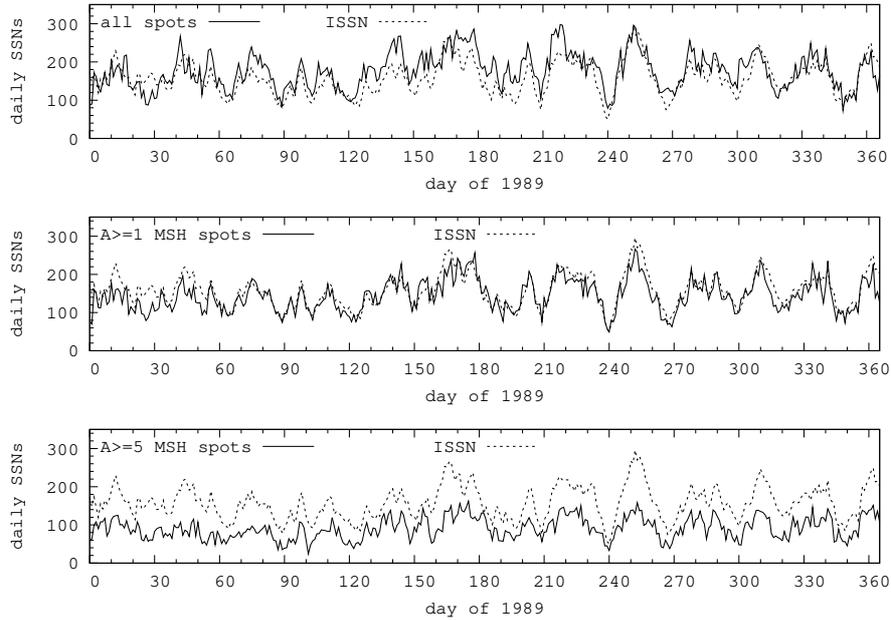}}
   
   \caption {Comparison of the ISSN (dashed lines) in 1989 with the Wolf number computed from the Debrecen Photoheliographic Data (continuous lines) by considering all spots (upper panel), omitting spots of 1 MSH (middle panel) and omitting all spots smaller than 5 MSH (bottom panel).}

   \label{ISSN_DPDWolf}
   \end{figure}

This observability problem raises the question whether we should omit the spots below an area limit. This could be a subjective intervention. Figure~\ref{ISSN_DPDWolf} shows the comparison of the daily values of ISSN and the Wolf number of Debrecen during 1989, the latter was computed by using the data of DPD with an observatory constant of unity. The figure shows the impact of omitting spots of size 1 MSH as well as all spots smaller than 5 MSH. The number of considered spots strongly affects the result.

\subsection {Methodological problems}
             \label{S-Wolfmethod}

\begin{figure}    
  \centerline{\includegraphics[width=0.3\textwidth,clip=,angle=-90]{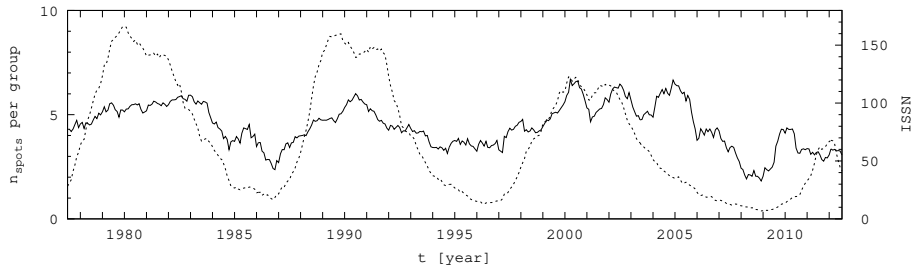}}
   
   \caption {Number of spots in groups at the phase of maximum area  within ±90º CMD in the DPD era by applying a 11 month smoothing. }

   \label{spotpergroup}
   \end{figure}

Another question may also be raised. How robust is the Wolf-definition? More specifically, what is the real weight of a group which has been chosen to be 10 in the definition? This can be checked by tracking the number of spots within the sunspot groups in the DPD era, between 1977 and 2014. see Figure~\ref{spotpergroup}. It can be seen that the weight of 10 is exaggerating on an average, although several groups may contain more than 10 spots.

As is well known the definition of the Group Sunspot Number \cite{Hoyt} contains a correction factor of 12.08 between the GSN and $R_{z} $. This value may also indicate a small mean number of spots per group in the historical observations contributing to the $R_{z} $. \inlinecite{Clette14} reports gradually diminishing GSN/ISSN ratio from 12.8 to 11.0 between cycles 19 and 24. If the number of spots per group plays the main role in this variation then Figure~\ref{spotpergroup} may mean that the variation of this ratio has also cyclic and mid-term components.

\begin{figure}    
   \centerline{\includegraphics[width=0.35\textwidth,clip=,angle=-90]{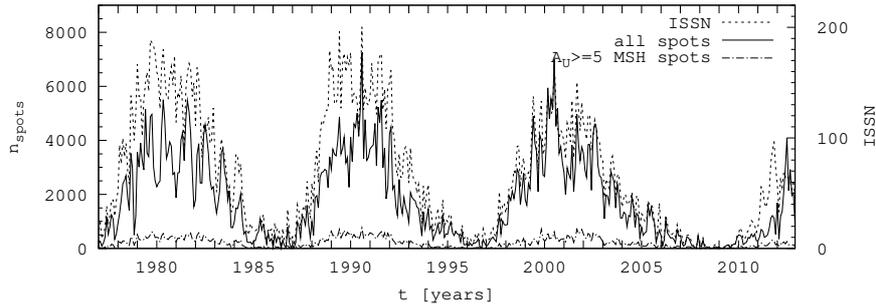}}
   
   \caption {Comparison of the ISSN (dashed line, scaled at right) in cycles 21--23 with the Wolf number computed from the DPD (scaled at left) by considering all spots (continuous line), and omitting spots smaller than 5 MSH (dashed dot line).}

   \label{haviISSN_DPD}
   \end{figure}

Figure~\ref{haviISSN_DPD} shows the comparison of the monthly mean ISSN and the number of all spots and spots larger than 5 MSH disregarding the groups during cycles 21--23 by using DPD data. These curves also show a strong dependence on the considered spot sizes.

There is a further methodological property which does not diminish the ambiguity, the redundant contribution of sunspot groups to the sunspot number. After the first appearance a group may exhibit rapid development and in the consecutive days it has varying contributions to the daily sunspot number. In such a way the variation of the sunspot number contains the mixed evolutional histories of several groups and these are smoothed out in the monthly ISSN values, although the contribution of an active region to the level of overall activity is only relevant once: at its maximum state.

\subsection {Problem of the physical meaning of the Wolf-definition}
                \label{S-Wolfmeaning}

Besides the observational constraints a conceptual remark can also be made on the Wolf-definition, namely, its physical meaning is ambiguous. Wolf considered the numbers of both the sunspot groups and the individual spots to be measures of the solar activity level. These two numbers, however, are signatures of two different physical mechanisms. 

The level of activity is characterised by the amount of the emerging magnetic flux, in other terms, the number of active regions and the flux carried to them from the toroidal field. In contrast, the number of individual spots within an active region is resulted by a fragmentation process during the flux emergence. \inlinecite{Fan} described a mechanism in which the plasma within the emerging flux ropes is carried towards the trailing part of the active region due to the Coriolis force and this results in the disintegration and dispersion of the spots of following polarity. In our previous paper \cite{Murakozy} we examined this fragmentation asymmetry on a large statistical sample and found that the leading part is generally more compact than the following one, it contains less and larger spots while the following part is more dispersed in the maximum state of the sunspot group development. 

We suggest that for the assessment of the solar activity level the amount of the emerged flux should be used. The sum of all emerged fluxes measured in the active regions during a cycle can be regarded as a proxy measure of the magnitude of the toroidal flux which is the source of the active regions.

\section {Total Amount of Emerged Magnetic Flux}
                \label{S-totalflux}

\begin{figure}    
   \centerline{\includegraphics[width=0.4\textwidth,clip=,angle=-90]{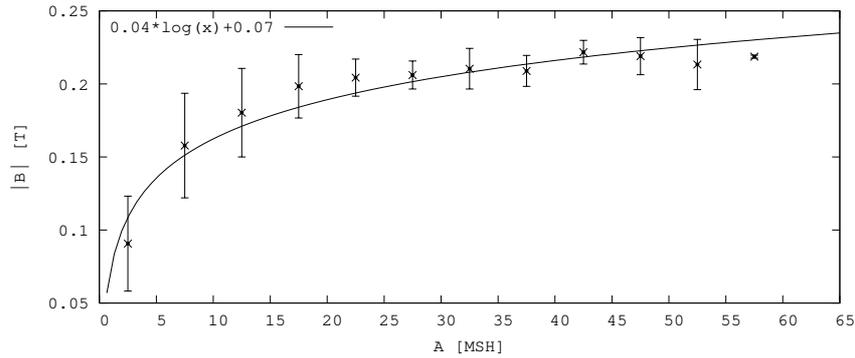}}
   
   \caption {Relationship between the umbral area and mean magnetic field by using SDD data based on SOHO/MDI observations.}

\label{A_B}
\end{figure}

The sunspot area datasets allow the amount of magnetic flux carried onto the solar surface to be calculated. This procedure needs a calibration function between the umbral area and the mean magnetic field enclosed in the umbra. This dependence is shown in Figure~\ref{A_B} which has been plotted by using the SDD data, i.e. SOHO/MDI observations. The sample consists of 44.780 sunspots, they were taken from the $10^{\circ}$ environment of the solar disc center in order to minimize the rate of magnetic field line deviations from the line-of-sight. The diagram does not distinguish between the magnetic polarities because they show the same relationship. The curve fitted to the points gives the area (A) -- magnetic field (B) relationship by the following formula:

\begin{equation}
      B = 0.04 \cdot logA + 0.07
\label{B_fA}
   \end{equation}

\noindent where $A$ is measured in Millionth of Solar Hemisphere (MSH) and B in Tesla.

By representing the B mean flux density with this function $f(A)$ the flux amount within the umbra  can be written as $ f(A) \cdot A $ . The total magnetic flux (TMF) carried by an active region can be represented in the following form:	

\begin{equation}
      TMF = K \cdot [ \Sigma  f(A_{i}) \cdot A_{i} ]_{LP} 
\label{TMF}
   \end{equation}

where
	$A_{i}$ : Area of i-th umbra (corrected for the geometrical foreshortening), 
	$f(A_{i}) = B_{i}$  the mean magnetic field of the i-th umbra, 
	$K$: ratio of the total and umbral fluxes in the active region, 
	LP: leading polarity.

The TMF should be computed in the maximum phase of a sunspot group because the most developed state shows the total amount of emerged flux, this also excludes the above mentioned multiple contribution of an active region to the final result. The restriction to the subgroup of leading polarity ensures that the flux is not taken into account twice, furthermore, the leading part is more reliable because of the above mentioned leading-following asymmetry. The $K$ correction factor takes into account the amount of small dispersed flux ropes belonging to the active region  out of the umbrae of high flux density. $K$ is regarded to be unity in the present study, it will be determined in a later work on a large sample for further refinement of the procedure. 

\begin{figure}    
   \centerline{\includegraphics[width=0.35\textwidth,clip=,angle=-90]{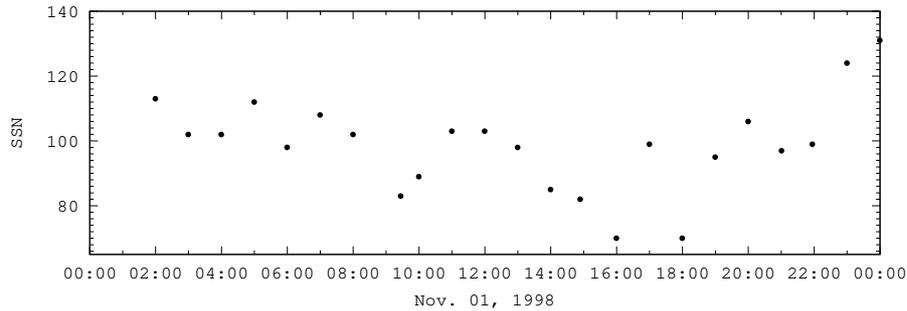}}
   
   \caption {Variation of the SSN (Wolf-number) during 24 hours in 1 November 1998.}

\label{981101}
\end{figure}

The advantage of the above parameter is that it avoids the necessity of correcting the LOS magnetic field measurements for all positions on the solar disc, it only considers the corrected umbral areas and the (1) calibration function. The correction for the geometrical foreshortening is more reliable than for the apparent variation of the magnetic field across the disc. Figure~\ref{981101} shows that the daily determination of the sunspot number may strongly depend on the time of observation even within 24 hours. The TMF helps to get rid of the role of sampling in the scatter of the result.

To produce a time series of the emerged flux a specific database has been made for the SOHO/MDI era by using the SDD sunspot data \cite{Gyori}. This database contains the list of all observed sunspot groups, the time of their maximum state and the list of all individual spots within the group with their corrected areas and magnetic polarities at the time of the maximum. This is the basis of computing their total magnetic fluxes, the above parameter TMF in (2). 

\begin{figure}    
   \centerline{\includegraphics[width=0.67\textwidth,clip=,angle=-90]{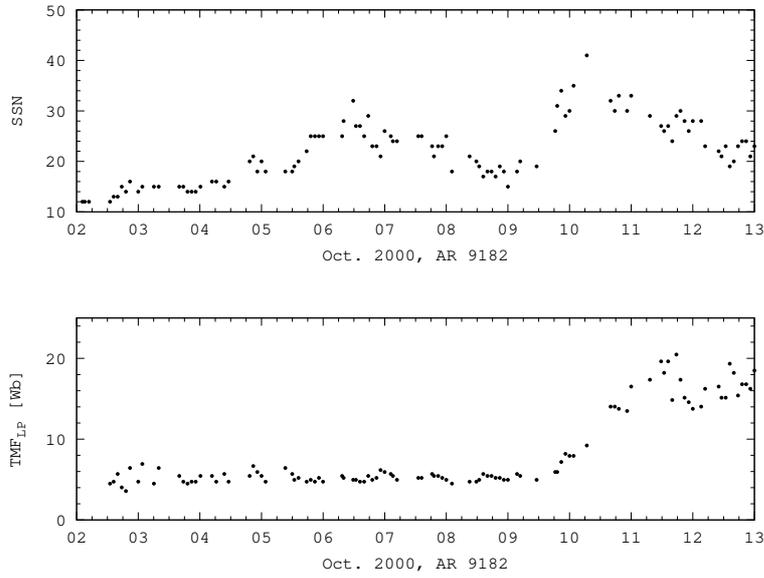}}
   
   \caption {Comparison of the development of sunspot group NOAA 9182 during its passage on the solar disc in October 2000 computed by the standard Wolf-definition (upper panel) and by TMF (lower panel).}

\label{9182AR}
\end{figure}

Figure~\ref{9182AR} compares the variation of a sunspot group during its passage through the solar disc computed by the Wolf-definition as if it was the only group on the disc (the calibration factor is 1) and the variation of its total magnetic flux. This latter is more reliable as can be checked visually in the html-presentation of the SDD catalogue (http://fenyi.solarobs.unideb.hu/SDD/2000/index.html).

It is reasonable to plot the variation of TMF monthly, because all active regions are considered only once. The monthly sums of TMFs are plotted in Figure~\ref{TMF} for the entire disc and for both hemispheres.

\begin{figure}    
  \centerline{\includegraphics[width=0.3\textwidth,clip=,angle=-90]{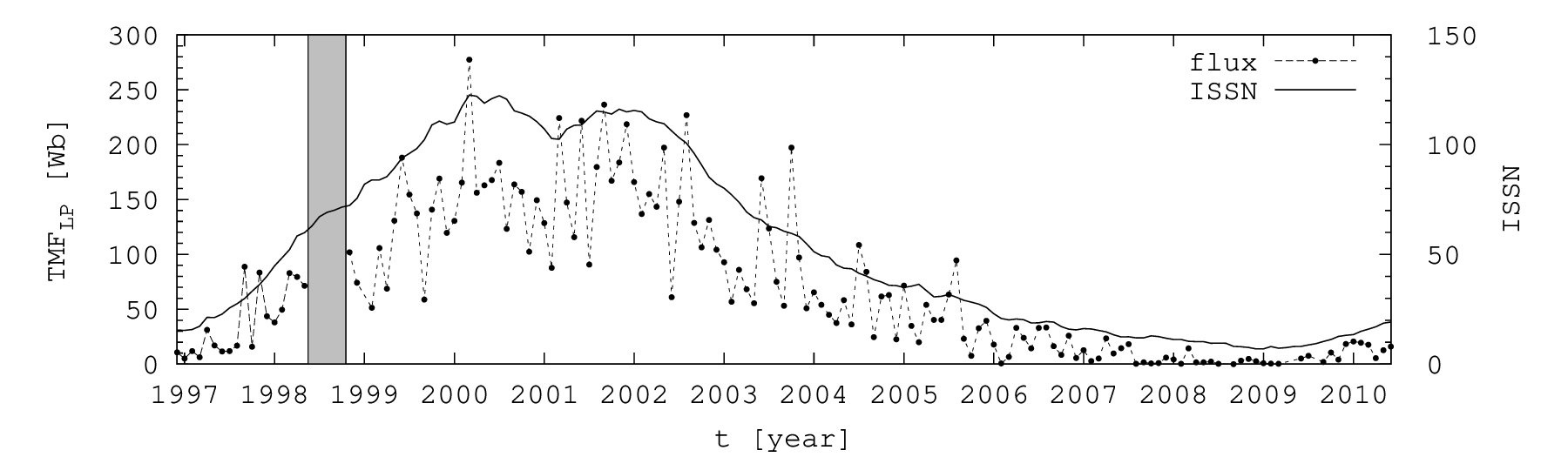}}
  \centerline{\includegraphics[width=0.3\textwidth,clip=,angle=-90]{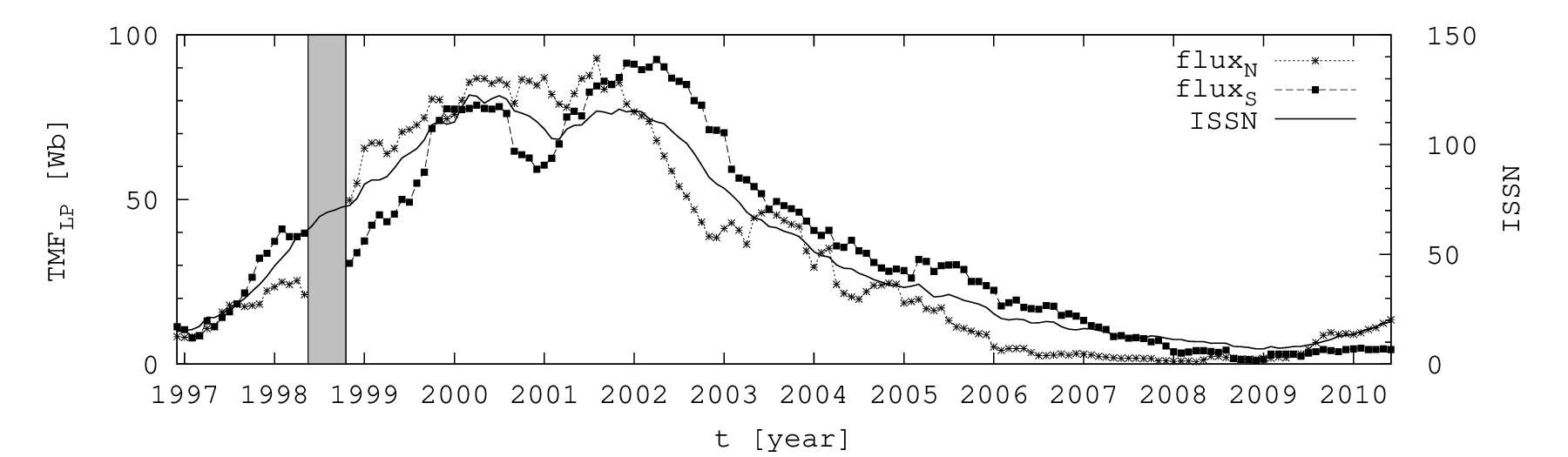}}
   
    \caption {Monthly sums of the total magnetic fluxes (TMF) in the sunspot groups for the entire disc (upper panel) and the hemispheric variations of TMF smoothed with a 11-month window (lower panel) in the SOHO/MDI era. The period of missing data is indicated with a vertical stripe. The time profile of ISSN is also smoothed with a 11-month window.}

\label{TMF}
\end{figure}

The TMF-contributions of all active regions can be summed up for an entire cycle and this can give a measure of the toroidal magnetic field strength in the given time interval. Between 1997 and 2010 the  sum of all TMFs was 4960.4 Wb on the northern hemisphere and 5695.5 Wb on the southern hemisphere.

Two remarks can be made. The maximum state can only be the maximum observed state which is not necessarily the true maximum because one can only follow the active regions until the western limb. All activity parameters are affected by this sampling restriction. 

The other remark concerns the (1) calibration function. We do not consider it a final version of the area-magnetic field relationship, but at present it can be regarded as an acceptable function following the recalibration of the MDI magnetograms \cite{Tran}. Upgrades of the calibration function may be motivated either by improved magnetic field measuring techniques to get rid of the saturation of magnetic data \cite{Liu}, or by intrinsic solar trends in the magnetic flux density within spots, see e.g.  \inlinecite{Livingston} and \inlinecite{Penn}. Any later refined functions will allow the TMF to be recomputed.

This procedure needs detailed sunspot area data but the currently available datasets do not yet allow to make long-term investigation of the TMF variations.  The most detailed materials are the SDD used here and the DPD also mentioned in Subsection~\ref{S-Wolfobs} covering cycles 21--24, the precision of their area data is about $10\% $. Cross calibrations of the available datasets are also necessary see. e.g. \inlinecite{Baranyi01}, \inlinecite{Balmaceda} and \inlinecite{Baranyi13}.

\section {Conclusions}

The presented proxy of the solar activity level, the total emerged magnetic flux (TMF) is physically more meaningful than the classic sunspot number defined by Wolf. This does not diminish the importance of the sunspot number that will remain the most important and indispensable long term proxy of the solar activity, especially in its recent form recalibrated by Clette et al (2014). We present here a possible approach for the determination of the TMF and its variation during cycle 23. 

Advantages of the suggested new parameter:

The TMF, the Total Magnetic Flux emerged within an active region (AR) see formula (2), is a physical quantity expressed in Weber in contrast to the sunspot number whose values can only be interpreted within the dataset itself.

The TMF of an AR has a well defined physical meaning, it is not contaminated with non-activity measures (sunspot fragmentation).

The TMF is also free from redundant contributions of consecutive days which also contaminates the sunspot number dataset with the evolution data of individual ARs. All ARs are considered once, at their maximum state but their reappearance has not been investigated here.

The contribution of small spots has less impact on the TMF than on the sunspot number because of the steeply diminishing flux density toward smaller umbrae (see Figure~\ref{A_B}).

Each AR has its own TMF contribution to the overall activity, their sum over a cycle may be a proxy measure of the strength of the toroidal field in the given cycle. In the interval 1997--2010 the  sum of all TMFs was 4960.4 Wb on the northern hemisphere and 5695.5 Wb on the southern hemisphere. The north-south difference can obviously not be characterized by this parameter alone, other features may exhibit different rates but it can be a relevant factor in the hemispheric asymmetry of the activity and presumably also in the asymmetry of the interplanetary field.

The following open questions remain for later studies:

The observational basis of this study is the SDD database, at present the only material allowing the determination of this proxy because no other datasets contain data for both the sunspot groups and sunspots along with their magnetic fields. This only allows the study of cycle 23.

The K correction factor in (2) will be targeted in a subsequent analysis to have a more reliable assessment of the entire emerged flux amount.

The procedure remains always open to recompute the flux amount if the $B=f(A)$ function in (1) will be upgraded by more precise magnetic field measurements.

The extension of the procedure for longer datasets with no magnetic data will need additional assumptions, primarily about the typical ratio of the leading and following parts in the most developed phase of the group, because the procedure is built on the leading part.

The aim is to have a parameter describing the solar activity which is not a mere number but it is a well defined physical quantity, the magnetic flux appearing at the solar surface. The cyclic sum of these quantities may characterize the total toroidal magnetic flux.

\begin{acks}

This work has been presented at the Sunspot Number Workshop held in Tucson, Arizona, January 2013, thanks are due to Edward W. Cliver for inviting us, J.M. and A.L.. The research leading to these results has received funding from the European Commission's Seventh Framework Programme (FP7/2007-2013) under the grant agreement eHeroes (project n° 284461, www.eheroes.eu).

\end{acks}

\end{article} 


\begin{thebibliography}{}

\bibitem[\protect\citeauthoryear{{Balmaceda} {\it et al.}}{2005}]{Balmaceda} Balmaceda, L., Solanki, S. K., Krivova, N.: 2005, \textit{Mem SAI} \textbf{76}, 929. 
ADS:\adsurl{2005MmSAI..76..929B}.
%doi:\doiurl{}.

\bibitem[\protect\citeauthoryear{{Baranyi} {\it et al.}}{2001}]{Baranyi01} Baranyi, T., Gy\H ori, L., Ludm\'any, A., Coffey, H.E.: 2001, \mnras{} \textbf{323}, 223. 
ADS:\adsurl{2001MNRAS.323..223B},
doi:\doiurl{10.1046/j.1365-8711.200104195.x}.

\bibitem[\protect\citeauthoryear{{Baranyi} {\it et al.}}{2013}]{Baranyi13} Baranyi, T., Kir\'aly, S., Coffey, H.E.: 2001, \mnras{} \textbf{434}, 1713. 
ADS:\adsurl{2013MNRAS.434.1713B},
doi:\doiurl{10.1093/mnras/stt1134}.

\bibitem[\protect\citeauthoryear{{Clette} {\it et al.}}{2007}]{Clette07} Clette, F., Berghmans, D., Vanlommel, P., Van der Linden, R.A.M., Koecklenbergh, A., Wauters, L.: 2007, \adv{} \textbf{40}, 919. 
ADS:\adsurl{2007AdSpR..40..919C},
doi:\doiurl{10.1016/j.asr2006.12.045}.

\bibitem[\protect\citeauthoryear{{Clette} {\it et al.}}{2014}]{Clette14} Clette, F., Svalgaard, L., Vaquero, J. M., Cliver, E. W.,: 2014, \ssr{} \textbf{186}, 35-103. 
ADS:\adsurl{2014SSRv..186..35C},
doi:\doiurl{10.1007/s11214-014-0074-2}.

\bibitem[\protect\citeauthoryear{{Fan} {\it et al.}}{1993}]{Fan} Fan, Y., Fisher, G.H., DeLuca, E.E.: 1993, \apj{} \textbf{405}, 390. 
ADS:\adsurl{1993ApJ...405..390F},
doi:\doiurl{10.1086/172370}.

\bibitem[\protect\citeauthoryear{{Gy\H ori} {\it et al.}}{2011}]{Gyori} Gy\H ori, L., Baranyi, T., Ludm\'any, A.: 2011, In: Choudhary, D.P., Strassmeier,K.G. (eds.), \textit{The Physics of Sun and Star Spots} \textbf{P-273}, IAU Symp., 403. \\ \href{http://fenyi.solarobs.unideb.hu/DPD/index.html}
ADS:\adsurl{2011IAUS..273..403G},
doi:\doiurl{10.1017/S174392131101564X}.

\bibitem[\protect\citeauthoryear{{Hoyt} and {Schatten}}{1998}]{Hoyt} Hoyt, D. V.,  Schatten, K. H.: 1998, \solphys{} \textbf{179}, 189. 
ADS:\adsurl{1998SoPh..179..189H},
doi:\doiurl{10.1023/A:1005007527816}.

\bibitem[\protect\citeauthoryear{{Livingston} and {Penn}}{2009}]{Livingston} Livingston, W., Penn, M.: 2009, EOS Trans. \textbf{90}, 257. 
ADS:\adsurl{2009EOSTr..90..257L},
doi:\doiurl{10.1029/2009EO300001}.

\bibitem[\protect\citeauthoryear{{Liu} {\it et al.}}{2007}]{Liu} Liu, Y., Norton, A., Scherrer. P.: 2007, \solphys{} \textbf{241}, 185. 
ADS:\adsurl{2007SoPh..241..185L},
doi:\doiurl{10.1007/s11207-007-0296-5}.

\bibitem[\protect\citeauthoryear{{Murak\"ozy} {\it et al.}}{2014}]{Murakozy} Murak\"ozy, J., Baranyi, T., Ludm\'any, A.: 2014, \solphys{} \textbf{289}, 563. 
ADS:\adsurl{2014SoPh..289..563M},
doi:\doiurl{10.1007/s11207-013-0416-3}.

\bibitem[\protect\citeauthoryear{{Penn} and {Livingston}}{2011}]{Penn} Penn, M., Livingston, W.: 2011, In: Choudhary, D.P., Strassmeier,K.G. (eds.), \textit{The Physics of Sun and Star Spots} \textbf{P-273}, IAU Symp., 126. 
ADS:\adsurl{2011IAUS..273..126P},
doi:\doiurl{10.1017/S1743921311015122 }.

%\bibitem[\protect\citeauthoryear{Royal Observatory, Greenwich}{1980}]{gpr}  Greenwich Photoheliographic Results, 1874-1976, in 103 volumes, 1980

\bibitem[\protect\citeauthoryear{{Tran} {\it et al.}}{2005}]{Tran} Tran, T., Bertello, L., Ulrich, R. K., Evans, S.: 2005, \apj{} \textbf{156}, 295. 
ADS:\adsurl{2005ApJS..156..295T},
doi:\doiurl{10.1086/326713}.

\end{thebibliography}
\end{document}